# Phenomenological Model of Longitudinal Spin Fluctuations in Itinerant Antiferromagnets

A. Solontsov[1,2,3] and V.P. Antropov[1]

[1]Ames Laboratory USDOE, Ames, IA 50011 USA
[2]A.A. Bochvar Institute for Inorganic Materials, 123060 Moscow, Russia
[3]State Center for Condensed Matter Physics, M. Zakharova str., 6/3, 155569, Moscow, Russia

**Abstract.** We present the phenomenological analysis of the spectrum of longitudinal spin fluctuations in isotropic itinerant electron antiferromagnets with account of spin anharmonicity giving rise to coupling of transverse and longitudinal normal modes. The spectrum consists of a quasielastic part forming a central peak or a dip, depending on temperature and the Landau relaxation rate. Effects of spin fluctuation coupling also give rise to an inelastic part of the spectrum which has a form of resonances or antiresonances near the magnon frequencies related to non-propagating longitudinal excitations.

Essentially, magnetic dynamics in itinerant electron magnets differs from that in the Heisenberg magnets due to the collective nature of itinerant electrons violating the conservation of magnetic moment amplitudes and giving rise to spin fluctuations (SF) which cannot be interpreted in terms of a simple precession of magnetic moments (see, e.g., [1]). Magnetic dynamics and SF in itinerant magnets play a decisive role in many technically important applications including Invar alloys [2], colossal magnetoresistive materials [3], high temperature superconductors [4], and newly discovered iron pnictide superconductors [5]. Whereas dynamics of transverse SF in itinerant magnets is well understood theoretically, the properties of longitudinal SF are still a puzzling point in the physics of magnetism (for a review see [6]). Contrary to dynamics of transverse fluctuations represented by the near-precession motion of magnetization density, linear dynamics of longitudinal SF in itinerant ferromagnets is thought to be dominated by Landau damping in the electron-hole continuum. Non-linear effects of the coupling of transverse and longitudinal SF result in additional channels of magnetic relaxation for both transverse and longitudinal SF. In itinerant ferromagnets, this leads to a novel mechanism of magnon damping caused by three-mode scattering processes of emission (absorption) of a longitudinal SF by a magnon [7], and it dominates over all other mechanisms of magnon damping specific for itinerant magnets [6,8]. In both Heisenberg and itinerant ferromagnets, coupling of longitudinal and transverse modes essentially affects quasielasic longitudinal SF and gives rise to inelastic longitudinal SF near the magnon frequencies [9-11], which is still under debate both theoretically [10-14] and experimentally (see, e.g., [15]).

Although some experimental evidence for the existence of quasielastic longitudinal SF of non-linear origin was presented by inelastic neutron scattering measurements [16,17], up to now, analysis of the mode-mode coupling effects on the spectrum of longitudinal SF in itinerant antiferromagnets is still lacking. Computer simulations of magnetic dynamics of Heisenberg antiferromagnets [14], as well as their analysis in the critical region in terms of renormalization-group techniques and model dynamical equations [18], however, could hardly shed light on the problem of longitudinal SF in itinerant antiferromagnets outside the critical region.

In this paper, we present the first phenomenological analysis of the spectrum of longitudinal SF strongly coupled to the transverse magnetic excitations in isotropic itinerant antiferromagnets basing on the mode-mode coupling ideas outside the critical region.

Let us now discuss the dynamical properties of an isotropic itinerant antiferromagnet in terms of the collective normal variables $m_\nu(k)$, i.e., the amplitudes of transverse ($\nu = t$) and longitudinal ($\nu = l$) SF, where $k = (\omega, \mathbf{k})$, $\omega$ and $\mathbf{k}$ are the frequency and wavevector, respectively, of SF measured from the center of the magnetic Brillouin zone. We shall consider SF with relatively long wavelengths and low frequencies, so that they are not dependent on the microscopic properties of a magnet and can be described within a phenomenological approach based on mode-mode coupling arguments. Below we consider a phenomenological model of itinerant antiferromagnets where ordered state can be described in terms of a single real order parameter, or staggered magnetization. This model was widely used to describe itinerant antiferromagnets with a commensurate spin density wave or magnetic sublattices [19]. Systems with incommensurate spin density waves like Cr and its alloys cannot be described by this simple model and are not discussed here.

Then linear magnetic dynamics of transverse SF is dominated by a precession motion characterized by weakly damped magnons, whereas dynamics of longitudinal SF has a purely relaxational character defined in the electronic ballistic regime by Landau damping in the electron-hole continuum. Then the transverse $\chi_t(k)$ and longitudinal $\chi_l(k)$ dynamical magnetic susceptibilities defining linear magnetic dynamics of an itinerant antiferromagnet can be taken in the form [19-21]

$$\chi_t(k) = \chi_t(\mathbf{k}) \frac{\omega_m^2(\mathbf{k})}{\omega_m^2(\mathbf{k}) - \omega^2 - 2i\omega\tau^{-1}}, \qquad (1)$$

and

$$\chi_l(k) = \chi_l(\mathbf{k}) \frac{\omega_{sf}(\mathbf{k})}{\omega_{sf}(\mathbf{k}) - i\omega}. \qquad (2)$$

The transverse dynamical susceptibility (Eq.(1)) has poles at the magnon frequencies $\omega = \pm\omega_m(\mathbf{k})$, where $\omega_m^2(\mathbf{k}) = \Delta^2 + (\mathbf{k}v_m)^2$, $\Delta$ is the spin wave gap, $v_m$ and $\tau^{-1}(\mathbf{k}) << \omega_m(\mathbf{k})$ are the velocity and inverse lifetime of weakly damped magnons, respectively. The longitudinal susceptibility (Eq.(2)) has a single imaginary pole at the SF frequency $\omega_{sf}(\mathbf{k}) = \Gamma_0(\mathbf{k})\chi_l^{-1}(\mathbf{k})$, where $\Gamma_0(\mathbf{k}) = \Gamma_0|\mathbf{k}|$ is the Landau relaxation rate of longitudinal SF, and is almost temperature independent, with $\Gamma_0 \sim v_F|\mathbf{k}|\chi_P$, where $v_F$ and $\chi_P$ are the Fermi velocity and Pauli susceptibility, respectively. The static susceptibilities in Eqs.(1) and (2) have the conventional form $\chi_{t,l}(\mathbf{k}) = \chi_{t,l}/[1 + (\mathbf{k}\xi_{t,l})^2]$, where $\xi_{t,l}$ are the transverse and longitudinal correlation lengths, and $\chi_{t,l}$ are staggered susceptibilities. The linear wavevector dependence of the scattering rate ($\Gamma_0(\mathbf{k})$) in itinerant antiferromagnets was recently confirmed by *ab initio* calculations of dynamical magnetic susceptibility in the iron arsenide family [22], and is different from the constraint $\Gamma_0(\mathbf{k}) = const(\mathbf{k})$ [19], which is usually applied in the SF theory of itinerant antiferromagnets. Here, we assume that Eqs.(1) and (2) have isotropic wavevector dependencies. Effects of spatial anisotropy do not principally change the results presented here (see [23]) and will be discussed elsewhere. It should be mentioned that to describe magnetic dynamics of more complicated antiferromagnets with incommensurate spin density waves one should account in Eq.(2) for additional poles related to longitudinal phason modes predicted for Cr and its alloys in [24].

We shall now consider mode-mode coupling effects in isotropic itinerant antiferromagnets. Among them, the most important ones include three-mode SF interactions, which were shown to play a dominating role in thermodynamics of itinerant antiferromagnets outside the critical region [19, 23]. To describe them, we shall use the simplest model based on the effective Ginzburg-Landau (GL) Hamiltonian for isotropic magnets [7,25]

$$\hat{H}_{eff} = W \sum_{k_1+k_2=k} m_l(k)[m_t(k_1)m_t^*(-k_2) + 3m_l(k_1)m_l(k_2)].$$  (3)

Here, $W = \gamma M$ is the matrix element, $M = M(T)$ is the staggered magnetization, $\gamma$ is the SF coupling constant, and $\sum_k = \sum_{\mathbf{k}} \int_{-\infty}^{+\infty} (d\omega/2\pi)$. Eq.(3) was widely used in the theory of critical phenomena [25] as the simplest model Hamiltonian where higher order terms in the order parameter contributing to the matrix element $W$ are neglected. We shall adopt this approach and similarly treat Eq.(3) as a model Hamiltonian which in our description is assumed to be valid down to low temperatures regardless of the low-temperature value of the staggered magnetization $M$. The equations of magnetic dynamics are then given by the following (Fourier transformed) time-dependent Ginzburg-Landau equations

$$\chi_t^{-1}(k)m_t(k) = - \sum_{k_1+k_2=k} W_{ltt}(k,k_1,k_2)m_l(k_1)m_t(k_2),$$  (4)

and

$$\chi_l^{-1}(k)m_l(k) = - \sum_{k_1+k_2=k} W_{ttl}(k,k_1,k_2)[m_t(k_1)m_t^*(-k_2) + W_{lll}(k,k_1,k_2)m_l(k_1)m_l(k_2)],$$  (5)

where the matrix elements describing SF couplings in the GL approach are assumed to be constants, $W_{ttl}(k,k_1,k_2) = W_{ltt}(k,k_1,k_2)/2 = W_{lll}(k,k_1,k_2)/3 = W = const(k,k_1,k_2)$. Here the right hand sides (r.h.s.) describe the effects of three-mode couplings of SF in isotropic antiferromagnets.

Provided the matrix elements in Eqs.(4) and (5) properly account for time and spatial dispersions the equations of magnetic dynamics have a general form following from the spin –invariance considerations for isotropic magnets [21]. Recently the matrix elements $\hat{W}$ were calculated microscopically within the Fermi liquid model for itinerant ferromagnets [26] where the GL values for them were confirmed for not very short wavelengths when $(\mathbf{k}\xi_l)^2 << 1$, i.e. not very close to the magnetic transition. A similar constraint holds for itinerant antiferromagnets. Below we shall assume that this inequality is satisfied.

Another limitation of the used here GL approach is related to the critical region defined by the Ginzburg criterion. Inside the critical region higher order mode-mode couplings may contribute to the longitudinal quasielastic spin fluctuation spectra, which we do not account for here.

Eqs. (1)-(5) define our mode-mode coupling model we use in this paper to describe longitudinal SF in itinerant antiferromagnets with account of their coupling to transverse excitations. In the past, we have used a similar model to analyze non-linear effects in the spectrum of transverse and longitudinal SF of itinerant ferromagnets [6,7,10,11].

As it follows from the dynamical Eqs. (4) and (5), coupling of SF may essentially influence the staggered magnetic susceptibilities in Eqs.(1)-(2). In the limit of weak spin anharmonicity, they should be replaced by [7]

$$\chi_t^{-1} \to \chi_t^{-1} + \gamma(2m_t^2 + 3m_l^2) , \ \chi_l^{-1} \to \chi_l^{-1} + \gamma\frac{\partial}{\partial M}(2m_t^2 + 3m_l^2) , \qquad (6)$$

where $m_\nu^2 = <\sum_k |m_\nu(k)|^2>$ are the average squared amplitudes of SF defining the temperature dependencies of magnetic susceptibilities which reproduce the well-known result of the thermodynamical approach to the theory of weakly anharmonic SF [19].

Most importantly, SF couplings in Eq.(3) open non-linear channels of magnetic relaxation due to three-mode scattering processes, which essentially modify magnetic dynamics. Here, we shall consider the effects of SF coupling on the spectrum of longitudinal SF. Besides the renormalization (Eq.(6)) of magnetic susceptibilities, SF couplings change the relaxation rate of the longitudinal SF, and should be replaced by

$$\frac{1}{\Gamma_0(\mathbf{k})} \to \frac{1}{\Gamma(k,T)} = \frac{1}{\Gamma_0(\mathbf{k})} + \frac{1}{\Gamma_n(k,T)} , \qquad (7)$$

where the second term in the r.h.s. describes the non-linear contribution to the relaxation rate of longitudinal SF. In the lowest order approximation accounting for three-mode coupling of longitudinal SF and magnons, the non-linear relaxation rate in Eq.(7) is given by

$$\Gamma_n^{-1}(k,T) = \frac{2\hbar}{\omega} \sum_{k'} |W|^2 \operatorname{Im} \chi_t(k') \operatorname{Im} \chi_t(k+k') \times \left( N_{\omega'} - N_{\omega'+\omega} \right), \qquad (8)$$

where $N_\omega = [\exp(\hbar\omega / k_B T) - 1]^{-1}$. It is necessary to emphasize that Eq.(8) does not depend on the spectrum of longitudinal SF and is a functional of the magnon spectrum only.

After substituting the transverse dynamical susceptibility (Eq.(1)) into Eq.(8), and assuming weak magnon damping, we have

$$\Gamma_n^{-1}(k,T) = \frac{\pi\hbar}{\omega} |W|^2 \sum_{\mathbf{k}} \sum_{\sigma=\pm 1} \frac{1}{\sigma\omega'} \omega'^2 \omega_m^2(\mathbf{k}+\sigma\mathbf{k}') \chi_t(\mathbf{k}') \operatorname{Im} \chi_t(\mathbf{k}+\sigma\mathbf{k}') \times$$

$$(9)$$

$$\delta[\omega_m^2(\mathbf{k}+\sigma\mathbf{k}') - (\omega+\sigma\omega')^2] \left( N_{\omega'} - N_{\omega'+\omega} \right)_{\omega'=\omega_m(\mathbf{k}')}.$$

As it follows from the integrand in the r.h.s., the frequencies of the scattered SF should satisfy the conservation laws

$$\omega_m(\mathbf{k}\pm\mathbf{k}') = |\omega \pm \omega_m(\mathbf{k}')| \qquad (10)$$

related to four types of scattering processes in isotropic antiferromagnets: emission (absorption) of a longitudinal SF by a magnon, and annihilation (creation) of two magnons giving rise to a longitudinal SF. It should be mentioned that the processes of annihilation (creation) of magnons are absent in ferromagnets where magnons have only one polarization.

Here, we shall assume that the spin wave gap and inverse correlation length of transverse SF are small compared to the maximum frequency $\omega_c$ and wavevector $k_c$ of magnons, i.e., $\Delta << \omega_c$, and $(k_c \xi_t)^2 >> 1$, so the following wavevector dependencies

$$\chi_t(\mathbf{k}) \approx \chi_t / (\mathbf{k}\xi_t)^2 \sim \mathbf{k}^{-2}, \quad \omega_m(\mathbf{k}) \approx v_m |\mathbf{k}| \sim |\mathbf{k}| \qquad (11)$$

hold in the major part of the Brillouin zone. With the use of Eq.(11), the integration in Eq.(9) yields the following explicit expression for the inverse non-linear relaxation rate:

$$\Gamma_n^{-1}(k,T) = \frac{1}{\Gamma_n |\mathbf{k}|} \vartheta(k) \frac{k_B T}{\hbar\omega} L(k,T). \qquad (12)$$

Here,

$$\Gamma_n^{-1} = \frac{\hbar^2}{16\pi} W^2 \chi_t^2(\mathbf{k}) \omega_m(\mathbf{k}) |\mathbf{k}|^3 \approx const(\mathbf{k}) \qquad (13)$$

is the relaxation constant which is wavevector independent, provided the equalities in Eq.(11) hold, and

$$L(k,T) = \ln \left| \frac{\exp\left[ -\dfrac{\hbar\left(\omega_m(\mathbf{k},T)+\omega\right)}{2k_BT} \right]-1}{\exp\left[ -\dfrac{\hbar\left(\omega_m(\mathbf{k},T)-\omega\right)}{2k_BT} \right]-1} \right|. \tag{14}$$

The function $\vartheta(k)$ accounts for the conservation laws resulting from Eq.(10): $\vartheta(k)=1$ for frequencies and wavevectors satisfying inequalities

$$-\omega_m(\mathbf{k})[2\omega_c - \omega_m(\mathbf{k})] < \omega(2\omega_c+\omega) < \omega_m(\mathbf{k})[2\omega_c + \omega_m(\mathbf{k})] \tag{15}$$

and $\vartheta(k)=0$ otherwise. For frequencies $\omega << \omega_c$, the phase space (Eq.(15)) reduces to $-(1-|\mathbf{k}|/2k_c)\omega_m(\mathbf{k}) < \omega < (1+|\mathbf{k}|/2k_c)\omega_m(\mathbf{k})$. In Eq.(14), we neglected the effects of weak magnon damping. We take them into account when necessary below. We shall also consider the spectrum of longitudinal SF within the phase volume (Eq.(15)) when $\vartheta(k)=1$, and omit $\vartheta(k)$ completely in further calculations.

Let us now discuss the spectrum of longitudinal SF with account of both mechanisms of magnetic relaxation: the linear mechanism due to Landau damping, and the non-linear mechanism caused by scattering of magnons on longitudinal SF. The spectrum is characterized by the intensity of longitudinal SF

$$I(k,T) = \frac{1}{\omega}\operatorname{Im}\chi_l(k), \tag{16}$$

and can be directly measured by inelastic neutron scattering. The longitudinal dynamical susceptibility accounting for coupling of longitudinal and transverse SF is given by Eqs. (2), (6), (7), and (12), and allow us to present the SF spectrum (Eq.(16)) in the explicit form

$$I(k,T) = I_0 \frac{\omega_m^2(\mathbf{k})[1+\xi(k,T)]}{\omega_{sf}^2(\mathbf{k})+[1+\xi(k,T)]^2\omega^2}, \tag{17}$$

where $I_0 = \chi_l(\mathbf{k})\omega_{sf}(\mathbf{k})/\omega_m^2(\mathbf{k})$. We also introduce the dimensionless parameter

$$\xi(k,T) = \frac{\Gamma_0(\mathbf{k})}{\Gamma_n(k,T)} = \frac{k_BT}{\hbar\omega}\frac{\Gamma_0}{\Gamma_n}L(k,T) \tag{18}$$

which describes non-linear effects of mode-mode coupling. As it follows from Eq.(18), mode-mode coupling vanishes in the low temperature limit. In the absence of coupling ($\xi = 0$), the spectrum of longitudinal SF (Eq.(17)) has a Lorentz form, and can be drastically altered by non-linear effects due to a rather complicated frequency dependence of the coupling parameter $\xi(k,T)$, as defined by Eq.(18), which also gives rise to a strong temperature dependence of the SF spectrum.

Here, we briefly discuss the analytical properties of $L(k,T)$ and $\xi(k,T)$. According to Eq.(14), $L(k,T)$ is an odd function of $\omega$, and $L(-k,T) = -L(k,T)$. In the low frequency limit

$$\omega << \omega_m(\mathbf{k}), \qquad (19)$$

$L(k,T)$ increases linearly with the frequency $L(k,T) \sim \omega$, and near the magnon frequencies $\omega \approx \pm \omega_m(\mathbf{k})$, it diverges logarithmically: $L(k,T) \sim \ln |\omega \mp \omega_m(\mathbf{k})|$. For relatively long wavelengths

$$\hbar \omega_m(\mathbf{k}) << k_B T, \qquad (20)$$

the expansion of Eq.(14) in powers of $\omega$ yields an almost linear dependence of $L(k,T)$ on the frequency:

$$L(k,T) \approx 2 \frac{\omega}{\omega_m(\mathbf{k})} [1 + \frac{1}{3}(\frac{\omega}{\omega_m(\mathbf{k})})^2 + ...]. \qquad (21)$$

Near the magnon frequencies, $L(k,T)$ is approximately given by

$$L(k,T) \approx \frac{1}{2} \ln \left| \frac{4\omega_m^2(\mathbf{k})}{[\omega_m(\mathbf{k}) \mp \omega]^2 + \tau^{-2}} \right|, \qquad (22)$$

where we account for weak magnon damping which prevents $L(k,T)$ from diverging.

For the coupling parameter (Eq.(18)) in the low frequency limit (Eq.(20)) with account of Eq.(21), we have

$$\xi(k,T) = \xi_0(\mathbf{k},T)[1 + \frac{1}{3}(\frac{\omega}{\omega_m(\mathbf{k})})^2 + ...], \qquad (23)$$

where the parameter

$$\xi_0(\mathbf{k},T) = 2 \frac{\Gamma_0}{\Gamma_n} \frac{k_B T}{\hbar \omega_m(\mathbf{k})} \qquad (24)$$

does not depend on the frequency and diverges in the long wavelength limit as $\left|\mathbf{k}\right|^{-1}$.

This parameter is strongly dependent on temperature: it disappears completely ($\xi_0(\mathbf{k},0)=0$) in the low-temperature limit, increases linearly at temperatures well below the Neel temperature where the magnon frequencies are almost temperature independent, and diverges ($\xi_0(\mathbf{k},T) \sim 1/\omega_m(\mathbf{k})$) as it approaches the magnetic phase transition. This divergence occurs provided the magnon frequencies vanish at the phase transition, as is believed to take place in most magnets. However, the vicinity of the magnetic transition cannot be described within used here DL approach.

Near the magnon frequencies $\omega = \pm\omega_m(\mathbf{k})$, the coupling parameter (Eq.(17)) has certain logarithmic anomalies

$$\xi(k,T) \approx \frac{1}{4}\xi_0 \ln\left|\frac{4\omega_m^2(\mathbf{k})}{[\omega_m(\mathbf{k}) \mp \omega]^2 + \tau^{-2}}\right|, \qquad (25)$$

where $\xi_0 = \xi_0(\mathbf{k},T)$ as is given by Eq.(24). This anomalous behavior of the coupling parameter results in drastic changes of the spectrum of longitudinal SF due to their coupling with transverse SF. First, effects of mode-mode coupling modify quasielastic SF, which are governed by the low frequency parameter (Eq. (23)) increase with a rise in temperature. Second, due to anomalies of the coupling parameter near the magnon frequencies (Eq.(25)), coupling of longitudinal and transverse SF gives rise to novel types of non-propagating longitudinal SF near the magnon frequencies.

To analyze the spectrum of quasielastic SF, we expand the intensity (Eq. (17)) in powers of $\omega$, and taking into account Eq. (23), we end up with

$$I(k,T) \approx I_0 \frac{1}{\eta^2(\mathbf{k})}\{1 + \xi_0 + \frac{\omega^2}{\omega_m^2(\mathbf{k})}[\frac{\xi_0}{3} - \frac{1}{\eta^2(\mathbf{k})}(1+\xi_0)^3] + ...\} . \qquad (26)$$

Here, $\eta(\mathbf{k}) = \omega_{sf}(\mathbf{k})/\omega_m(\mathbf{k})$ is a dimensionless parameter nearly independent of the wavevector in the long wavelength limit. The sign of the coefficient in the term $\sim \omega^2$ in the r.h.s. of Eq. (26) characterizes the type of quasielastic SF. For

$$3\frac{[1+\xi_0(\mathbf{k},T)]^3}{\xi_0(\mathbf{k},T)} > \eta^2(\mathbf{k}), \qquad (27)$$

the coefficient in the $\sim \omega^2$ term in Eq.(26) is negative, and is related to a quasielastic peak in the SF spectrum. When the inequality (Eq.(27)) is violated,

$$3\frac{(1+\xi_0(\mathbf{k},T))^3}{\xi_0(\mathbf{k},T)} < \eta^2(\mathbf{k}),\tag{28}$$

quasielastic SF are characterized by a dip at $\omega = 0$, which separates inelastic satellites near the magnon frequencies. As it follows from Eq.(27), the dip does not appear when

$$\eta(\mathbf{k}) < 9/2.\tag{29}$$

According to Eq.(17), the spectrum of longitudinal SF takes the simple Lorentz form describing a central quasielastic peak in the limits of weak and strong mode-mode coupling. For weak mode-mode coupling (i.e., at low temperatures), where

$$\xi_0(\mathbf{k},T) << 1,\tag{30}$$

the spectrum of longitudinal SF (Eq.(17)) has the conventional form

$$I(k,T) = \chi_l(\mathbf{k})\frac{\omega_{sf}(\mathbf{k})}{\omega_{sf}^2(\mathbf{k}) + \omega^2},\tag{31}$$

and describes a quasielastic peak of a linear nature with the half width at half maximum (HWHM)

$$\Lambda(\mathbf{k}) = \omega_{sf}(\mathbf{k}) \sim |\mathbf{k}|,\tag{32}$$

which is linearly dependent on the wavevector in the long wavelength limit ($(\xi_l \mathbf{k})^2 << 1$) due to Landau damping. With an increase in temperature, the coupling parameter (Eq.(24)) increases, and in the strong coupling limit where

$$\xi_0(\mathbf{k},T) >> 1,\tag{33}$$

as it follows from Eq.(17), the spectrum of longitudinal SF has the Lorentz form again, but of a purely non-linear nature:

$$I(k,T) = \chi_l(\mathbf{k})\frac{\omega_{sf}(\mathbf{k})/\xi_0(\mathbf{k},T)}{[\omega_{sf}(\mathbf{k})/\xi_0(\mathbf{k},T)]^2 + \omega^2}.\tag{34}$$

Following from Eq.(33), the non-linear quasielastic peak in Eq.(34) has the intensity $I(k,T)_{\omega=0} \sim \chi_l(\mathbf{k})T/\omega_m(\mathbf{k})\omega_{sf}(\mathbf{k})$, which increases rapidly with a rise in temperature. The HWHM

$$\Lambda(\mathbf{k}) = \omega_{sf}(\mathbf{k})/\xi_0(\mathbf{k},T) \sim \omega_m(\mathbf{k})\omega_{sf}(\mathbf{k})/T \sim \mathbf{k}^2\tag{35}$$

of the non-linear peak described by Eq.(34) has a quadratic wavevector dependence in the long wavelength limit ($(\xi_l \mathbf{k})^2 \ll 1$), and is in contrast to the linear dependence in Eq.(31) characterizing the width of the linear peak. This may be a clue for separating linear versus non-linear quasielastic SF in neutron scattering investigations of various antiferromagnetic systems.

The evolution from linear to non-linear quasielastic peaks described by Eqs. (31) and (34) with an increase of mode-mode coupling and temperature can take place according to two different scenarios depending on the ratio $\eta(\mathbf{k}) = \omega_{sf}(\mathbf{k})/\omega_m(\mathbf{k})$. The first scenario is realized in itinerant antiferromagnets satisfying the inequality in Eq.(29), where the width of the linear quasielastic peak $\omega_{sf}(\mathbf{k})$ is less or comparable to the magnon frequency. Therefore, quasielastic longitudinal SF at low temperatures where non-linear effects are negligible are well resolved from the transverse propagating magnons. In this scenario, the linear quasielastic peak arising at low temperatures gradually transforms into the high temperature non-linear peak.

Another scenario describing the temperature dependence of the spectrum of quasielastic SF is realized in itinerant antiferromagnets with

$$\eta(\mathbf{k}) > 9/2, \qquad (36)$$

where the width of the linear quasielastic peak $\omega_{sf}(\mathbf{k})$ is larger than the magnon frequency, and therefore, quasielastic longitudinal SF at low temperatures are merged with transverse propagating magnons. As temperature rises, the low temperature linear regime of SF characterized by the quasielastic Lorentz peak given by Eq.(31) is changed by a dip developing at $\omega = 0$ in the temperature range defined by Eq.(28). When the rising temperature eventually exceeds the range defined by Eq.(28), the dip transforms into a quasielastic peak (Eq.(34)) of a purely non-linear origin. The latter scenario of the temperature-dependent behavior of SF therefore differs from the former one due to the development of a quasielastic dip in the intermediate temperature range given by Eq.(28).

Besides quasielastic SF, coupling of longitudinal and transverse modes results in inelastic longitudinal SF developing near the magnon frequencies $\omega = \pm\omega_m(\mathbf{k})$. According to Eqs.(17), (18), and (22) at finite temperatures, two satellite peaks arise, and their intensity increases rapidly with a rise in temperature. To compare intensities of the inelastic and quasielastic peaks, one should estimate the ratio

$$\frac{I(k,T)_{\omega=0}}{I(k,T)_{\omega=\omega_m(\mathbf{k})}} = \frac{1+\xi_0}{1+\xi_m}\left[1+\frac{\omega_m^2(\mathbf{k})}{\omega_{sf}^2(\mathbf{k})}\left(1+\xi_m\right)^2\right],\tag{37}$$

where $\xi_m = \xi(k,T)_{\omega=\omega_m(\mathbf{k})} \approx (\xi_0/2)\ln[2\omega_m(\mathbf{k})\tau]$. As temperature and SF coupling (Eq.(24)) rise in antiferromagnets satisfying the constraint of Eq.(29), the intensity of the quasielastic peak increases monotonically when compared to the inelastic peaks, according to Eq.(37). This is accompanied by a gradual change in the nature of the quasielastic peak from linear to a non-linear one. In these antiferromagnets, quasielastic peaks dominate over inelastic ones over the entire temperature range below the Neel temperature.

In antiferromagnets where Eq.(36) holds, the intensity of inelastic peaks increases with a rise in temperature, and they dominate over the central quasielastic peak. At temperatures defined by $\xi_0(\mathbf{k},T) \sim 1/3\eta^2(\mathbf{k})$, the quasielastic peak transfers into a dip between the inelastic peaks at $\omega = \pm\omega_m(\mathbf{k})$. As temperature continues to rise until the parameter $\xi_0(\mathbf{k},T) \sim \eta^2(\mathbf{k})/\sqrt{3}$, the dip vice versa transforms into a quasielastic peak of a purely non-linear origin.

The inelastic part of the spectrum of longitudinal SF resulting from resonances of the r.h.s. of Eq.(17) near the magnon frequencies can have a fine structure. As temperature and SF coupling rise, each of the inelastic resonances developing near $\omega = \pm\omega_m(\mathbf{k})$ grows in intensity, and at temperatures defined by

$$\xi_m(\mathbf{k},T) \sim \eta(\mathbf{k},T),\tag{38}$$

they are transformed into antiresonances accompanied by two satellite peaks. This fine structure of the inelastic spectrum of SF arising at temperatures satisfying Eq.(38) exists up to the Neel temperature.

Now we can summarize the analysis of the temperature dependencies of the quasielastic and inelastic parts of the spectrum of longitudinal SF caused by coupling of longitudinal and transverse SF.

The first scenario takes place in itinerant antiferromagnets with low $\eta$ satisfying Eq.(29). In the low temperature limit where SF coupling satisfies Eq.(30), the spectrum of longitudinal SF has a relatively narrow quasielastic peak related to linear paramagnon-like excitations arising due to the Landau damping, and its width (Eq.(32)) is linearly dependent on the wavevector. Coupling to the transverse SF gives rise to two inelastic resonances at the magnon frequencies $\omega = \pm\omega_m(\mathbf{k})$, and are well separated from the quasielastic peak due to the inequality in Eq.(29). When temperature rises, the intensity of the resonances grows due to the increase of SF coupling, and at temperatures defined by Eq.(38), the resonances are converted into antiresonances accompanied by two satellites each. Up to this temperature, the spectrum of longitudinal SF generally has a three-peak structure consisting of a central quasielastic peak and two inelastic peaks near the magnon frequencies with the following fine structure: antiresonances at $\omega = \pm\omega_m(\mathbf{k})$ with two satellite each. As the temperature rises even higher, the intensity of the central quasielastic peak grows faster than the inelastic ones, and it dominates over them. At high temperatures where SF coupling (Eq.(33)) is strong, the spectrum of longitudinal SF has a purely non-linear nature, and its shape is formed by a central peak dominating over antiresonances. The width of the central peak (Eq.(35)) depends quadratically on the wavevector (in contrast to the linear dependence at low temperatures).

Another, more complicated scenario takes place in itinerant antiferromagnets with relatively high $\eta$ satisfying Eq.(36), where at low temperatures, the spectrum consists of a broad quasielastic peak due to Landau damping with relatively narrow inelastic resonances at the magnon frequencies superimposed on the central peak. Unlike the first scenario, where both the central peak and inelastic resonances grew with rises in temperature, the increase of resonances is more pronounced here, and at temperatures defined by Eq.(28), the central peak separating the inelastic resonances transforms into a dip. With a further rise in temperature, the intensity of inelastic SF increases while the resonances are transformed into antiresonances. At the relatively high temperatures defined by Eq.(27), the central dip transforms back into a peak, and grows rapidly with increases in temperature until it dominates over the antiresonances. Further development of the SF spectrum takes place in accordance with the first scenario discussed above. The scenarios for the temperature dependence of the spectrum of longitudinal SF in itinerant antiferromagnets defined by Eqs.(29) and (36) are illustrated by Figs. 1 , and show the evolution of the spectrum of longitudinal SF calculated from Eq.(17) with the variation of temperature and (or) spin anharmonicity.

Direct experimental evidence for existing quasielastic longitudinal SF has been presented [16] for the cubic isotropic antiferromagnet $RbMnF_3$ with the Neel temperature 83K using inelastic neutron scattering data together with the polarization analysis. The authors discovered a quasielastic component of the spectrum with longitudinal polarization which had the shape of a central peak increasing rapidly while approaching the Neel temperature and decreasing with the increase of the wavevector. The behavior of this central longitudinal peak clearly indicates its non-linear nature and can be well described using the quasielastic spectrum (Eq.(34)).

Quasielastic longitudinal SF have been also observed [17] in the isotropic itinerant antiferromagnet UN with the Neel temperature 53K using inelastic neutron scattering. The intensity of the peaks of quasielastic SF was shown to increase rapidly with a rise in temperature and presented direct evidence for non-linear nature of longitudinal SF observed in UN.

Perspective candidates for the observation of non-linear effects of SF coupling are the iron pnictide itinerant antiferromagnets, and are parent compounds for the recently discovered superconductors [5]. They possess high frequency magnons with the energies up to 100meV [5,22], which according to Eq.(13), may give rise to strong non-linear magnetic relaxation affecting the spectrum of longitudinal SF. Besides, inelastic neutron scattering measurements of antiferromagnetic iron pnictides [22] discovered the presence of significant Landau damping – ultimate manifestation of the electron itinerancy. In addition, our density functional calculations also revealed strong Landau damping of magnons [22] and comparable values of transverse and longitudinal components of the static magnetic susceptibility [27,28], which strongly supports their itinerant character. All these findings make iron pnictides an ideal system for the application of the approach described above.

In summary, we presented the first analysis of the spectrum of longitudinal SF in itinerant electron antiferromagnets taking into account coupling of longitudinal and transverse SF. At moderate temperatures, SF coupling essentially affects quasielastic SF and gives rise to inelastic non-propagating longitudinal excitations near the magnon energies. At higher temperatures, the spectrum of longitudinal SF is dominated by a rapidly growing quasielastic central peak of a non-linear nature. Our results form the basis for the understanding of longitudinal SF in itinerant antiferromagnets at finite temperatures, when spin anharmonicity may play an important role mixing normal longitudinal and transverse modes.

It should be also emphasized that the presented here results open a new fundamental problem of itinerant magnetism. Namely, up to now it was thought that SF thermodynamics in itinerant magnets is well described by the self-consistent renormalized (SCR) theory of Moriya [19] in the weak coupling limit and by soft mode (SM) theory of Solontsov and Wagner [6,23], both assuming that SF are caused by the linear (Landau) mechanism of magnetic relaxation. On the other hand, the presented above results clearly show that at elevated temperatures SF may be caused by a non-linear mechanism of magnetic relaxation resulting in a more complicated spectrum of longitudinal SF. This finding should lead to a reexamination of the results of the SCR and SM theories of SF.

Work at the Ames Laboratory was supported by Department of Energy-Basic Energy Sciences, under Contract No. DE-AC02-07CH11358. AS would also acknowledge the support of ROSATOM and Russian Foundation for Basic Research (grant No 06-02-17291).

### References.

**Figure captions**

Fig. 1: Spectrum of longitudinal SF as a function of temperature $t = k_B T / \hbar \omega_m(\mathbf{k})$ and coupling parameter $\xi_0(\mathbf{k}, T)$, calculated from Eq.(17) for $\omega_m(\mathbf{k})\tau = 0.001$ and $\Gamma_0 / \Gamma_n = 10$. Curves 1 (blue), 2 (dirty yellow), 3 (red), and 4 (green) are calculated for t=0.05 ($\xi_0 = 1.0$), t=0.2 ($\xi_0 = 4.0$), t=0.3 ($\xi_0 = 6.0$), and t=0.5 ($\xi_0 = 10.0$). (a) Temperature evolution of the spectrum in the scenario characterized by Eq.(29) with $\omega_{sf}^2(\mathbf{k}) / \omega_m^2(\mathbf{k}) = 10$. The intensity of the central quasielastic peak increases rapidly with the increase in temperature and spin anharmonicity, accompanied by resonances (curve 1), or antiresonances (curves 2, 3, and 4) related to non-propagating longitudinal spin fluctuations near the magnon frequencies $\omega = \pm \omega_m(\mathbf{k})$. (b) Temperature dependence of the spectrum in the scenario defined by Eq.(36) with $\omega_{sf}^2(\mathbf{k}) / \omega_m^2(\mathbf{k}) = 30$. With the increase in temperature the wide central peak (curve 1) transforms into a dip (curve 2), which transforms back into a central peak (curve 3) and grows in intensity (curves 4). Quasielastic spin fluctuations are surrounded by resonances (curve 1), or antiresonances (curves 2, 3, and 4) characterizing non-propagating longitudinal spin fluctuations developing near $\omega = \pm \omega_m(\mathbf{k})$.

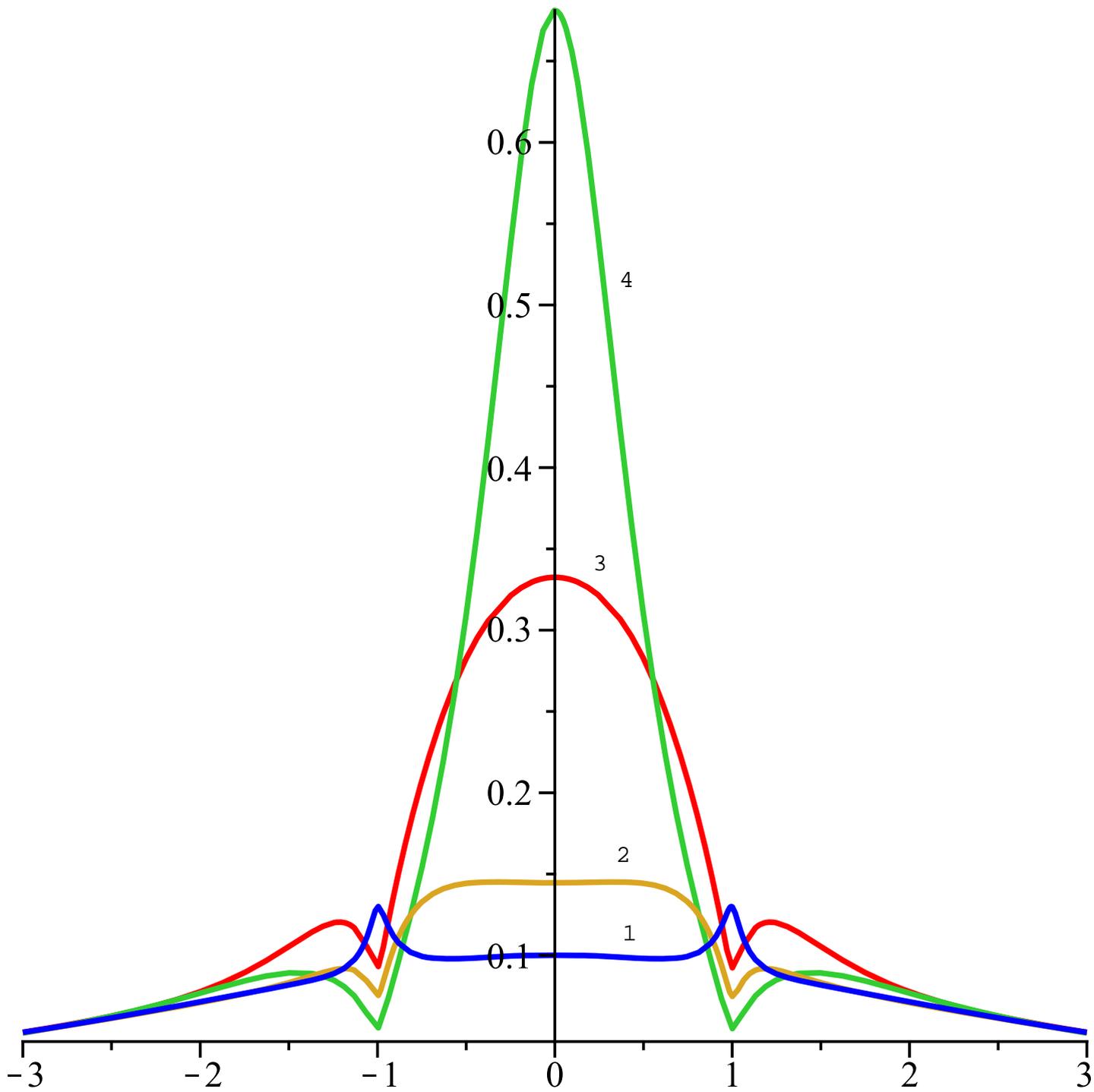

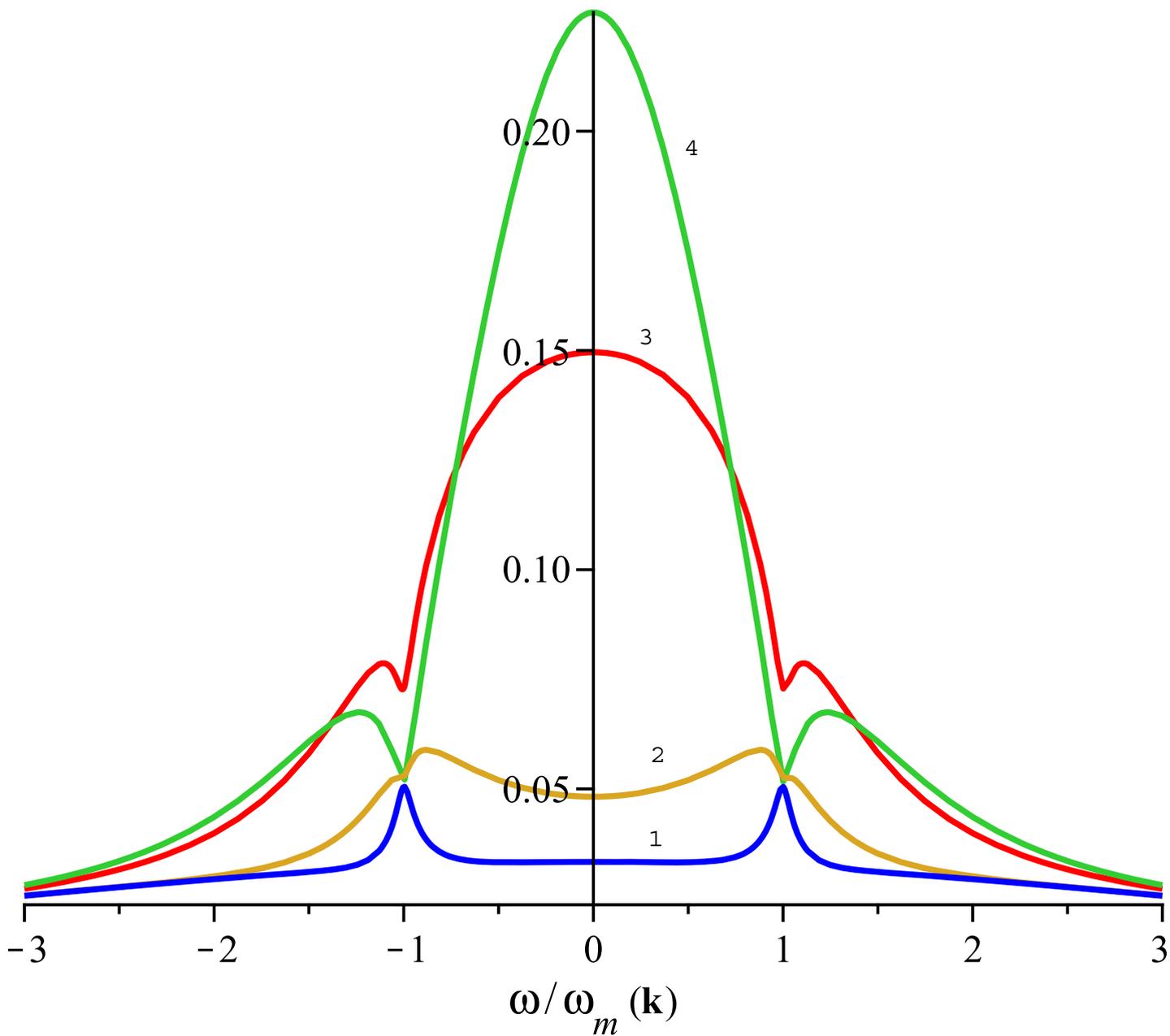